\documentclass[journal,twoside]{IEEEtran}

\usepackage{amsmath,amssymb,amsfonts}

\usepackage{algorithm}
\usepackage{algorithmic}

\usepackage{graphicx}
\usepackage[caption=false]{subfig}

\usepackage{tabularx}
\usepackage{adjustbox}

\usepackage{cite}

\usepackage{textcomp}
\usepackage{times}

\usepackage[hidelinks]{hyperref}

\def\BibTeX{{\rm B\kern-.05em{\sc i\kern-.025em b}\kern-.08em
    T\kern-.1667em\lower.7ex\hbox{E}\kern-.125emX}}

\title{FetalSleepNet: A Transfer Learning Framework with Spectral Equalisation Domain Adaptation for Fetal Sleep Stage Classification}
\author{%
Weitao~Tang, \IEEEmembership{Student Member,~IEEE},
Johann~Vargas-Calixto, \IEEEmembership{Member,~IEEE}, \\Nasim~Katebi, \IEEEmembership{Member,~IEEE}\thanks{J. Vargas-Calixto, N. Katebi, and G. D. Clifford are with the Department of Biomedical Informatics, Emory University, Atlanta, USA.}, 
Nhi Tran, Sharmony~B.~Kelly\thanks{Nhi Tran, R. Galinsky and S. B. Kelly are with the Ritchie Centre, Hudson Institute of Medical Research, Melbourne, Australia.}, 
Gari~D.~Clifford\thanks{G. D. Clifford is also with the Department of Biomedical Engineering, Georgia Institute of Technology, Atlanta, USA.}, \IEEEmembership{Fellow,~IEEE},
Robert~Galinsky\thanks{R. Galinsky is also with the Department of Obstetrics and Gynaecology, Monash University, Melbourne, Australia.}, 
Faezeh~Marzbanrad\thanks{W. Tang and F. Marzbanrad are with the Department of Electrical and Computer Systems Engineering, Monash University, Melbourne, Australia.}, \IEEEmembership{Senior Member,~IEEE}
\thanks{This work was supported in part by the NIH (R01HD110480), Google.org AI for the Global Goals Impact Challenge, and NHMRC (1124493\&1164954). N.K. is partially supported by a PREHS-SEED award (K12ES033593).}
}
\begin{document}
\maketitle

\begin{abstract}
Introduction: This study presents FetalSleepNet, the first published deep learning approach 
%and label-efficient 
to classifying sleep states from the ovine electroencephalogram (EEG). Fetal EEG is complex to acquire and difficult and laborious to interpret consistently.   
%which serve as important indicators of neurophysiological development. 
However, accurate sleep stage classification may aid in the early detection of abnormal brain maturation associated with pregnancy complications (e.g. hypoxia or intrauterine growth restriction). 

%We evaluated several deep learning strategies for fetal sleep stage classification using electroencephalogram 

Methods: EEG electrodes were secured onto the ovine dura over the parietal cortices of 24 late-gestation fetal sheep. 
%To the best of our knowledge, FetalSleepNet is the first deep learning framework specifically developed for automated sleep staging from the fetal EEG. 
A lightweight deep neural network originally developed for adult EEG sleep staging was trained on the ovine EEG using transfer learning from adult EEG. A spectral equalisation–based domain adaptation strategy was used to reduce cross-domain mismatch. 

Results: We demonstrated that while direct transfer performed poorly, full fine-tuning combined with spectral equalisation achieved the best overall performance (accuracy: 86.6\%, macro F1-score: 62.5), outperforming baseline models. %convolutional neural network with long short-term memory (CNN+LSTM) trained on handcrafted EEG features (accuracy: 82.6\%, macro F1-score: 58.3), and the same architecture trained on raw EEG (accuracy: 76.4\%, macro F1-score: 54.9). 
% Such metrics can elucidate the evolution of sleep in utero, and facilitate frequent, cost-effective monitoring of fetal neural development and early detection of pathological conditions. 

Conclusions: To the best of our knowledge, FetalSleepNet is the first deep learning framework specifically developed for 
automated sleep staging from the fetal EEG. 
Beyond the laboratory, the EEG-based sleep stage classifier functions as a label engine, enabling large-scale weak/semi-supervised labeling and distillation to facilitate training on less invasive signals that can be acquired in the clinic, such as Doppler Ultrasound or electrocardiogram data. FetalSleepNet’s lightweight design makes it well suited for deployment in low-power, real-time, and wearable fetal monitoring systems. 

\end{abstract}

\begin{IEEEkeywords}
Deep learning, Electroencephalography, Real-time systems, Sleep, Transfer learning
\end{IEEEkeywords}

\section{Introduction}
% \IEEEPARstart{S}{leep} state patterns reflect fetal neurophysiological function and development~\cite{schneider2008human}, %,zizzo2020fetal,mulkey2018critical,samjeed2024fetal,hoyer2017monitoring,roffwarg1966ontogenetic}, 
%  and are clinically relevant for detecting abnormal neurodevelopment, which may result from conditions such as chronic hypoxia and intrauterine growth restriction~\cite{shaw2018altered,arduini1989behavioural,van1985behavioural}.
\IEEEPARstart{S}{leep} state patterns reflect fetal neurophysiological function and development~\cite{schneider2008human}, and are clinically relevant for detecting abnormal neurodevelopment, which may result from conditions such as chronic hypoxia, infection or hypertensive disorders of pregnancy (HDP)~\cite{shaw2018altered,arduini1989behavioural,van1985behavioural}.
% koos1987fetal 
The fetal sheep is a well-established preclinical model for in-utero brain function and sleep–state development, exhibiting late-gestation rapid eye movement (REM) and non-rapid eye movement (NREM) organization and hypothalamic sleep–wake circuitry akin to human fetuses~\cite{schwab2000investigation,lee2002prostaglandin}. % schwab2006time,tournier2022physiological
 This model enables chronic instrumentation, allowing repeated access to maternal and fetal blood and cerebrospinal fluid, thus facilitating clinically relevant studies of oxygen and nutrient transport, hemodynamics, and neurodevelopment~\cite{back2012instrumented}. The model has contributed significantly to prenatal medicine, including the development of therapies for premature birth, hypoxia and fetal growth complications~\cite{back2012instrumented,10784962}.

Fetal sheep, similar to humans, present canonical sleep states, which include REM, NREM, and an intermediate state~\cite{nijhuis1982there,szeto1985prenatal,tang2025fetalsleepcrossspeciesreview}. However, sleep state definitions in fetal research remain a topic of debate, particularly concerning the existence of true fetal wakefulness~\cite{tang2025fetalsleepcrossspeciesreview}. Transitional or ambiguous states are especially difficult to delineate, with physiological studies showing considerable variability in their characterization and interpretation~\cite{tang2025fetalsleepcrossspeciesreview}. This lack of consensus reflects both the complexity of fetal neurodevelopment and the methodological limitations of current assessment frameworks.

Traditional manual annotation of fetal sleep using electroencephalogram (EEG) is labor-intensive and inherently subjective. Visual scoring suffers from inconsistencies in state labeling, especially during transitions, and lacks scalability for large datasets. These limitations underscore the urgent need for automated, objective classification tools that can ensure reproducibility and cross-study comparability. In particular, label-efficient and domain-adaptive deep learning approaches are essential for biomedical sensing scenarios like fetal EEG, where collecting large, well-annotated datasets is costly and time-consuming.

Despite growing interest in deep learning for sleep staging, applying these methods to fetal EEG (fEEG) poses unique challenges. Compared with adult EEG, fEEG suffers from a scarcity of large labeled datasets and substantial domain differences, including electrode placement, signal amplitude, frequency content, and the neurodevelopmental maturation of sleep states. In fetal sheep (term $\approx$ 21 weeks), organized electrocorticographic sleep–wake cycling with REM and NREM organization emerges at approximately 16–17 weeks of gestation (late gestation) and consolidates toward term~\cite{szeto1985prenatal}; in humans, unsynchronized cycles first appear at approximately 32 weeks but coherent behavioural states are not reliably established until 38–40 weeks~\cite{nijhuis1982there}; by childhood and adulthood, NREM–REM cycles show stable~\cite{feinberg1979systematic}, curvilinear trends across the night—highlighting developmental shifts that limit the direct transfer of adult sleep models to automatic fetal sleep staging.

Automated fetal sleep staging could deliver major translational benefits by enabling state-specific analyses of physiological signals such as heart rate variability, maternal–fetal heart rate interactions \cite{marzbanrad2015quantifying}, and fetal movement. Sleep state is not currently considered in antenatal monitoring, yet incorporating it could improve the accuracy of fetal wellbeing assessments by allowing clinicians to cross-validate electronic monitoring results against active and quiet behavioural states. Robust EEG-based labels position the EEG sleep stager as a label engine, enabling weak/semi-supervised pretraining and distillation of proxy stagers from non-invasive signals such as Doppler ultrasound and fetal electrocardiogram (fECG). As a brain-based bridge toward human translation, fetal magnetoencephalography (fMEG) can support transfer from fetal sheep and neonatal EEG to a human fetal EEG sleep stager, ultimately contributing to the establishment of normative growth curves, early detection of abnormalities such as hypoxia, intrauterine growth restriction or infection, and improved assessments of fetal viability, sensory development, and risk of HDP.

% To our knowledge, this is the first study to achieve automatic classification of fetal sleep states using EEG. We introduce FetalSleepNet, the first deep learning framework specifically designed for fetal EEG (fEEG)-based sleep staging, adapted from TinySleepNet~\cite{supratak2020tinysleepnet}, a compact CNN originally trained on adult human sleep EEG from the Sleep-EDF dataset~\cite{kemp2000analysis,goldberger2000physiobank}. By tailoring this architecture to fetal EEG, FetalSleepNet provides a scalable and objective solution for fetal sleep classification. This work establishes a foundation for advancing our understanding of fetal neurodevelopment and highlights the potential of fEEG-based monitoring for early detection of abnormal conditions and future clinical translation.
To address these challenges, we introduce FetalSleepNet, the first deep learning framework specifically designed for fEEG-based sleep staging. Our approach provides a scalable and objective solution for fetal sleep classification. This work lays the foundation for advancing our understanding of fetal neurodevelopment and highlights the potential of fEEG-based monitoring for early detection of abnormal conditions, while providing a scalable label engine for proxy stagers and a translational pathway targeting improved fetal and maternal health risk prediction.

\begin{table*}[htbp]
\centering
\caption{Comparison of FBS Classification Studies}
\label{tab:combined_fbs_comparison}
\renewcommand{\arraystretch}{1.6}
\large
\resizebox{\textwidth}{!}{%
\begin{tabular}{|p{2.5cm}|p{2.5cm}|p{1.3cm}|p{2.5cm}|p{3cm}|p{3.8cm}|p{4.8cm}|p{6.4cm}|}
\hline
\textbf{Study} & \textbf{Population} & \textbf{\# Individuals} & \textbf{Gestational Age} & \textbf{Signals Used} & \textbf{Method} & \textbf{States Identified} & \textbf{Performance} \\
\hline
Vairavan et al. (2016)~\cite{vairavan_computer-aided_2016} & Human fetuses & 39 & 30--38 weeks & fMCG (HR + Actogram) & Rule-based thresholds + ROC optimization & 
\begin{tabular}[t]{@{}l@{}}$<$36 wks: 1F vs. 2F \\ $\geq$36 wks: 1F vs. 2F\end{tabular} & 
\begin{tabular}[t]{@{}l@{}}$<$36 wks: ICC = 0.88 (1F), 0.65 (2F) \\ $\geq$36 wks: ICC = 0.88 (1F), 0.41 (2F) \\ AUC = 0.99 (both)\end{tabular} \\
\hline

Semeia et al. (2022)~\cite{semeia_evaluation_2022} & Human fetuses & 52 & 27--39 weeks & fMCG (HRV + Actogram) & Rule-based thresholds + ROC optimization & 
\begin{tabular}[t]{@{}l@{}}$<$32 wks: Active vs. Passive \\ $\geq$32 wks: 1F vs. 2F\end{tabular} & 
\begin{tabular}[t]{@{}l@{}}$<$32 wks: AUC $\approx$ 1.0 (HRV), \\0.80--0.83 (Actogram) \\ $\geq$32 wks: AUC $\approx$ 1.0 (HRV), \\0.86--0.87 (Actogram)\end{tabular} \\
\hline

Myers et al. (1993)~\cite{myers_quantitative_1993} & Fetal baboons & 3 & 20--22 weeks & EEG (frontal + parietal) & Rule-Based (K-means preprocessing) & 1F vs. 2F & 
\begin{tabular}[t]{@{}l@{}}Expert agreement: 87.1\% (Overall)\\ 79.7\% (1F), 91.3\% (2F)\end{tabular} \\
\hline

Grieve et al. (1994)~\cite{grieve1994behavioral} & Fetal baboons & 3 & 23--25 weeks & EEG, EOG, ECG & Rule-Based (K-means preprocessing) & 1F vs. 2F & 
\begin{tabular}[t]{@{}l@{}}Expert agreement: 81.5\% (Overall) \\ 83.7\% (1F), 79.4\% (2F)\end{tabular} \\
\hline

Samjeed (2022)~\cite{samjeed2022classification} & Human fetuses & 105 & 20--40 weeks & Non-invasive fetal ECG (NI-fECG) & 1D CNN & 1F vs. 2F & 
\begin{tabular}[t]{@{}l@{}}F1: 80.2\% (1F), 69.5\% (2F) \\ Accuracy: 76\% \\ Sensitivity: 72.7\% (1F), 82.6\% (2F)\end{tabular} \\
\hline

Subitoni (2022)~\cite{subitoni2022hidden} & Human fetuses & 115 & 27--39 weeks (grouped: early/mid/late) & FHR & HMM + CNN (Hybrid) & 1F vs. 2F & 
\begin{tabular}[t]{@{}l@{}}HMM+CNN: \\ F1: 87.87\%, Balanced Acc: 88.37\% \\ HMM only: \\ F1: 77.73\%, Balanced Acc: 83.30\%\end{tabular} \\
\hline
\end{tabular}
}
\end{table*}

\subsection*{Statement of Contributions}
The main contributions of this work are as follows:
\begin{itemize}
    \item We propose FetalSleepNet, the first deep learning framework for automatic classification of sleep states in fetal sheep using EEG. 
    % It adapts TinySleepNet with architectural modifications for dual-channel inputs and is trained using transfer learning to adapt to the fetal sleep domain.
    
    \item We propose a lightweight and effective preprocessing technique—spectral equalisation—to reduce the frequency-domain mismatch between adult and fEEG, enabling more efficient transfer learning and faster convergence.

%    \item We further conduct a systematic comparison of common transfer learning strategies—frozen CNN, partial fine-tuning, and full fine-tuning—in the context of fetal sleep staging. Although these approaches are standard in other EEG domains, their relative effectiveness has not been evaluated for this task. Our results show that, when applied together with the spectral equalisation preprocessing described earlier, full fine-tuning consistently achieves the best performance, setting a new benchmark on our fetal sleep dataset.

    % \item We benchmark FetalSleepNet against conventional CNN+LSTM models trained directly on fEEG, using both raw signals and handcrafted features. %Our results show that transfer learning from adult EEG outperforms fetal-only models, particularly in REM and NREM classification, highlighting its utility in data-scarce scenarios.

    \item We publicly release the full implementation of FetalSleepNet, including pretrained weights on spectrally-equalised adult EEG and fine-tuned weights on fetal sheep EEG. This enables direct application to fetal sheep data for sleep classification, as well as transfer learning to human or other species. The code will be made publicly available upon acceptance (with a DOI-registered release via Zenodo), and is currently hosted on GitHub~\cite{Tang_FetalSleepNet_A_Transfer_2025}.

    %\item We frame the sleep stager as a label engine, enabling large-scale weak/semi-supervised labeling to pretrain Doppler/ECG sleep stagers on broader clinical datasets.

    %\item We outline a translational pathway targeting HDP risk prediction, leveraging fMEG as a brain-based bridge for transferring from sheep/neonatal EEG to a human fetal EEG stager.

\end{itemize}
\section{Related Works}
Fetal behavioral state (FBS) classification has been explored using a variety of physiological signals and modeling approaches in both human and animal studies. Table~\ref{tab:combined_fbs_comparison}, reproduced from our prior review~\cite{tang2025fetalsleepcrossspeciesreview}, provides a comparative overview of representative FBS classification studies across modalities, species, and algorithms. Note that in all of these studies, the two-state classification of 1F and 2F is used to represent quiet sleep (analogous to NREM) and active sleep (analogous to REM), respectively.

Traditional methods rely on rule-based criteria derived from fetal heart rate variability (FHRV), actogram-derived movement, or EEG spectral features. For instance, Vairavan et al.~\cite{vairavan_computer-aided_2016} and Semeia et al.~\cite{semeia_evaluation_2022} proposed fetal magnetocardiography (fMCG)-based systems that achieved high agreement with expert scoring using HRV-derived thresholds, though performance declined for active sleep and across gestational stages.

In animal studies, early work in fetal baboons used unsupervised K-means clustering on EEG and multimodal signals. Myers et al.~\cite{myers_quantitative_1993} introduced the EEG-ratio as a marker of behavioral state, while Grieve et al.~\cite{grieve1994behavioral} extended this with electrooculography (EOG) and RR variability to define binary sleep states with ~81–87\% agreement to expert annotations. These methods provided valuable physiological insight but were constrained by small sample sizes and absence of an established ground truth, and therefore could not leverage the scalability and pattern-recognition advantages that deep learning-based automation can offer.

More recently, machine learning models have been applied to fECG or FHR signals. Samjeed et al.~\cite{samjeed2022classification} trained a 1D convolutional neural network (1D-CNN) on non-invasive fECG recordings, achieving 76\% accuracy, while Subitoni et al.~\cite{subitoni2022hidden} proposed a hybrid hidden markov model (HMM)-CNN model using FHR signals, reaching a Macro F1-score of 87.87\%. However, these studies rely on indirect proxies of brain activity, and expert scoring inconsistencies remain a challenge. Most human fetal studies rely on non-invasive signals such as fECG, fMCG, or HRV to estimate sleep states, due to the inability to acquire fEEG in utero. These signals provide only indirect markers of central nervous system activity. In contrast, fetal sheep models enable direct, invasive cortical EEG recording, allowing objective classification of neurophysiological sleep states. Such models also provide a foundation for cross-species and cross-modality translation, ultimately guiding future human fetal studies.

To our knowledge, this study presents the first automated method for classifying sleep states in fetal sheep using EEG, leveraging animal data as a biological reference to support future studies using non-invasive techniques in human fetuses. Furthermore, FetalSleepNet is the first deep learning framework specifically developed for fEEG-based sleep staging, enabling end-to-end learning directly from raw EEG without manual feature engineering.

% In addition, our framework explicitly positions the fetal EEG sleep stager as a label engine, providing reliable annotations to pretrain proxy sleep stagers from non-invasive signals (e.g., Doppler, fECG) via weak/semi-supervised learning. As a translational bridge, fMEG can further enable transfer from sheep and neonatal EEG to a human fetal EEG sleep stager, ultimately supporting large-scale studies and downstream applications such as risk prediction of HDP.

% \newpage
\section{METHODS}

\subsection{Data Collection}
\subsubsection{Fetal Sheep}
All procedures were approved by the Hudson Institute of Medical Research Animal Ethics committee and were conducted in accordance with the ARRIVE guidelines~\cite{percie2020arrive} and the National Health and Medical Research Council Code of Practice for the Care and Use of Animals for Scientific Purposes (Eighth Edition).

Twenty four pregnant Border-Leicester ewes carrying singletons at 118-119 days gestation underwent sterile surgery for fetal instrumentation. EEG was measured from the left and right cerebral hemispheres using two pairs of electroencephalogram (EEG) electrodes (AS633-7SSF; Cooner Wire, Chatsworth, CA, USA) placed through burr holes onto the dura over the parasagittal parietal cortex (10 and 20 mm anterior to bregma, and 10 mm lateral)~\cite{tran2025prophylactic}.

Continuous EEG recordings at 121 days of gestational age (dGA) were captured and continued until the end of the study period. All signals were collected using commercial hardware (Powerlab, ADInstruments, Australia) and recorded continuously using LabChart Pro software (v8.1.16; ADInstruments, Australia), sampled at 400 samples/s. The raw EEG signals were saved in LabChart’s native binary format (.adicht) for downstream processing and analysis. The recording duration lasted 2 weeks.

\subsubsection{Adult}

We used the Sleep Cassette subset of the Sleep-EDF Expanded dataset, publicly available from PhysioNet~\cite{kemp2000analysis,goldberger2000physiobank}. The dataset contains 153 overnight polysomnographic recordings from 78 healthy adult participants aged 25–101 years (37 males and 41
females), collected between 1987 and 1991. Participants were recorded at home over two consecutive nights using a portable cassette-tape EEG system. Each recording includes EEG (Fpz–Cz and Pz–Oz), EOG, submental Electromyography (EMG), and event markers. EEG and EOG signals were sampled at 100~Hz.

Corresponding hypnograms were annotated by trained technicians using the 1968 Rechtschaffen and Kales (R\&K) criteria~\cite{rechtschaffen1968}, with stages Awake, REM, Stage N1–N3, Movement, and Unknown. %We used the EEG channels and sleep labels to pretrain a deep learning model on adult sleep before transferring to fetal data.

\subsection{Data Annotation}
To support sleep state classification, we annotated sleep states in 24 fetal sheep recordings based on visual inspection of EEG signals, guided by established electrophysiological criteria~\cite{tang2025fetalsleepcrossspeciesreview,rao2009behavioural}. Each 12-hour recording was segmented into three sleep states: NREM, REM, and Intermediate.

Sleep staging was primarily based on Left EEG (LEEG) and Right EEG (REEG) signals, as well as their corresponding Spectral Edge Frequencies (L SEF and R SEF) ~\cite{thaler2000real}. Spectral Edge Frequency (SEF90) was calculated using a 32,768-point Blackman-windowed FFT, with a 90\% cumulative power threshold. The zero-frequency component was removed to avoid low-frequency bias. This was computed for both LEEG and REEG channels to aid in state annotation. Depending on signal quality, either the left or the right channel served as the primary reference.

NREM sleep was defined as high-voltage, low-frequency EEG activity with low SEF values in both hemispheres, persisting for at least 3 min. REM sleep was defined as low-voltage, high-frequency EEG with elevated SEF values, also persisting for at least 3 min~\cite{rao2009behavioural,galinsky2017connexin,galinsky2023magnesium}. 

Segments that did not meet either of these definitions were grouped into the \textit{Intermediate} state category, which included:
\begin{itemize}
    \item Sleep cycles with duration shorter than 3 min;
    \item Segments with mismatched EEG amplitude and SEF frequency;
    \item Transitional EEG features showing mixed characteristics between REM and NREM.
\end{itemize}

This approach allowed us to account for ambiguous or unstable sleep states without discarding data.

\begin{figure}[htbp]
    \centering
    \includegraphics[width=\linewidth]{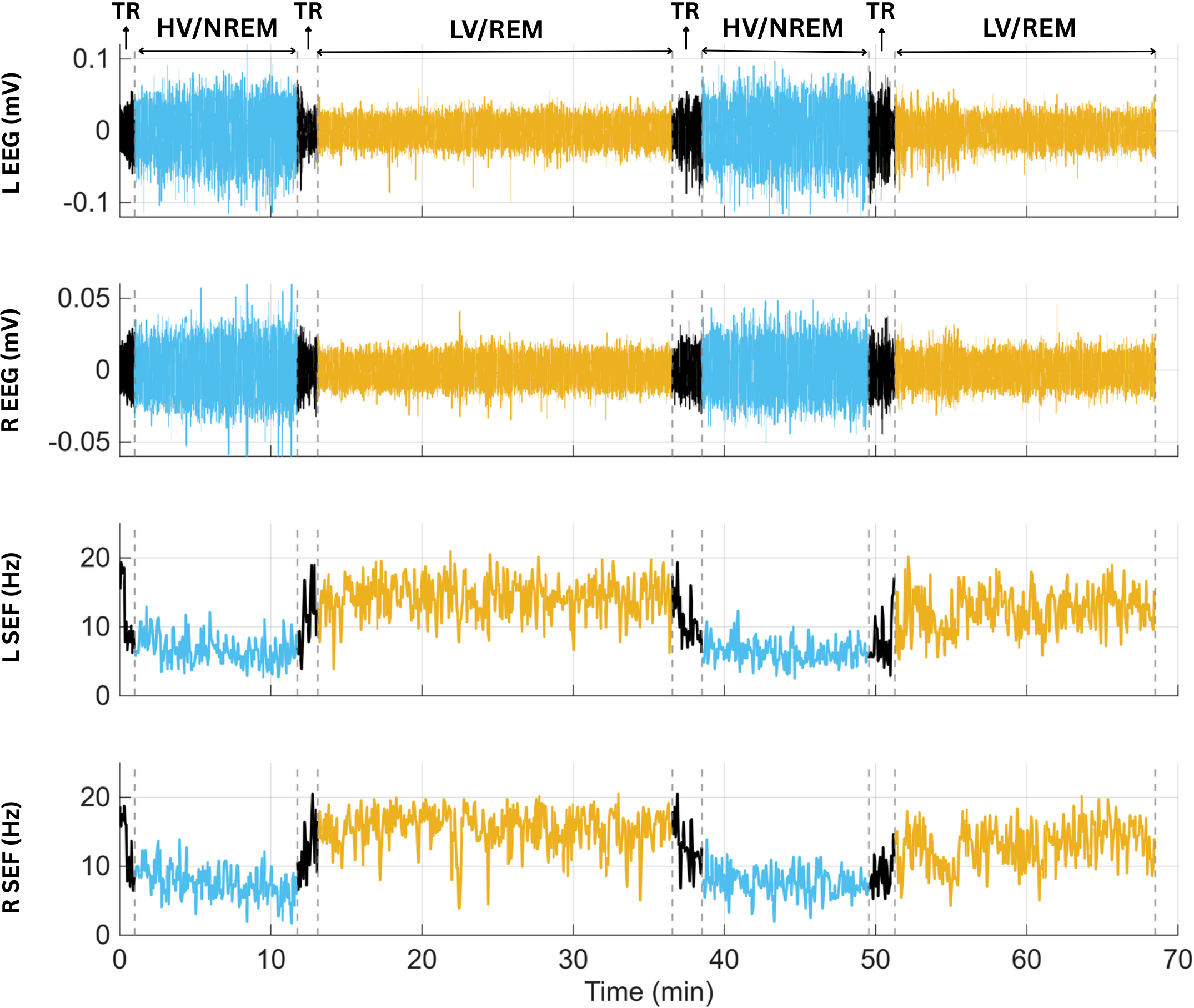}
    \caption{Representative physiological signals illustrating sleep states in fetal sheep~\cite{tang2025fetalsleepcrossspeciesreview}. NREM sleep is marked by high-voltage (HV), low-frequency EEG patterns recorded from both hemispheres (LEEG and REEG), whereas REM sleep is characterized by low-voltage (LV), high-frequency EEG activity. TR represents a transitional state between REM and NREM, typically exhibiting mixed EEG features that do not clearly conform to either state. In addition to raw EEG signals, spectral edge frequencies (L SEF and R SEF, computed using the 90\% threshold) provide complementary frequency-domain features that aid in both visual interpretation and manual annotation of sleep states. %All signals were obtained from chronically instrumented fetal sheep using bilateral EEG recordings, enabling continuous in utero monitoring of neural activity.
    }
    \label{fig:label_presentation}
\end{figure}

Fig.~\ref{fig:label_presentation} presents representative signal segments from the LEEG and REEG channels, along with their corresponding L SEF and R SEF values. These examples illustrate three canonical sleep patterns: REM, NREM, and a Transition state (TR). Note that the Transition state is a subtype within the broader Intermediate category, which also includes short-duration or ambiguous segments that cannot be clearly classified as REM or NREM.

Annotations were performed by W.T. under the supervision of R.G., who provided expert neurophysiological guidance. Both authors independently reviewed the entire dataset, and any discrepancies were discussed and resolved through consensus to ensure consistency.

\subsection{Pre-Processing}
We performed signal pre-processing and feature extraction in MATLAB, focusing primarily on EEG signals recorded from fetal sheep. Each file contained continuous EEG recordings acquired at a sampling rate of 400 Hz from two parietal cortex channels (LEEG and REEG), each lasting approximately 12 hours, with expert-annotated sleep states (REM, NREM, and Intermediate). To ensure high data quality, we excluded segments where sleep stage annotations could not be confidently determined due to indistinct EEG patterns, or labeling disagreements between annotators. Such ambiguous segments typically occurred during state transitions or in the presence of transient artifacts (e.g., motion or electrode noise). Although the exact proportion varied across recordings, these segments represented only a small fraction of the total data and their removal did not alter the class distribution.

First, independent component analysis (ICA)-based artifact removal was applied to reduce non-neural noise using EEGLAB~\cite{delorme2004eeglab}. The EEG data were restricted to the union of labeled regions, and ICA was performed only once per recording. Component rejection was guided by the ICLabel algorithm, and only components classified as non-brain (e.g., muscle, eye, line noise) with less than 90\% probability were removed. Following artifact removal, a finite impulse response (FIR) bandpass filter (1–22 Hz) was applied to retain the physiologically relevant frequency content of fEEG while suppressing slow drifts and high-frequency noise.

Preprocessed signals were segmented into overlapping 30-second windows with a 15-second step size, consistent with the standard epoch length defined in adult sleep scoring guidelines such as the 1968 Rechtschaffen and Kales (R\&K) criteria~\cite{rechtschaffen1968} and the AASM manual~\cite{iber_aasm_2007}. Each window was assigned a sleep state label using a majority-voting scheme based on temporal overlap with annotated intervals.

From each window, we extracted a comprehensive set of handcrafted features from both LEEG and REEG channels. These included:

\begin{itemize}
\item \textbf{Time-domain features}: mean, standard deviation, peak-to-peak amplitude, and zero-crossing count~\cite{hamida2015eeg}.
\item \textbf{Hjorth parameters}: activity, mobility, and complexity—quantifying the signal’s amplitude, frequency spread, and structural complexity~\cite{hjorth1970eeg}.
\item \textbf{Frequency-domain features}: Absolute and relative power (dB) were computed within four frequency bands—Delta (1–3.9 Hz), Theta (4–7.9 Hz), Alpha (8–12.9 Hz), and Beta (13–22 Hz)—based on power spectral density (PSD) estimated using Welch's method~\cite{kelly2023progressive,welch2003use}.
\item \textbf{Nonlinear features}: Detrended Fluctuation Analysis (DFA) and Petrosian Fractal Dimension (PFD), which capture signal complexity and self-similarity~\cite{10784962}.
\item \textbf{Inter-channel coherence}: Theta-band (4--7.9 Hz) coherence between LEEG and REEG channels, indicative of interhemispheric phase synchronization, which has been linked to functional coupling during sleep and memory processes~\cite{fell2003rhinal}.
\end{itemize}

Absolute EEG amplitude thresholds for sleep state classification—such as low-voltage REM ($<$50~$\mu$V), Intermediate-voltage (50--100~$\mu$V), and high-voltage NREM (100--200~$\mu$V)—were originally proposed by Rao et al.~\cite{rao2009behavioural}. We observed similar amplitude-based patterns in our fetal sheep recordings; however, consistent application of these thresholds was hindered by inter-subject variability in EEG amplitude, which arose from both physiological differences and technical factors such as electrode placement and signal quality. To mitigate this issue and ensure reliable interpretation of amplitude-based state distinctions, we applied z-score normalization on a per-subject, per-channel basis prior to model training. This step also improves model generalization by standardizing input distributions. Notably, the Intermediate state lies between NREM and REM in both behavioral characteristics and amplitude, making it particularly sensitive to baseline voltage shifts, which further supports the need for normalization.

%In parallel, we prepared a raw EEG dataset suitable for deep learning models. 
To ensure consistency with the Sleep Cassette dataset, the LEEG and REEG signals were converted from millivolts ($m$V) to $\mu$V by scaling the signals $\times 1000$. The filtered EEG signals were first downsampled from 400 Hz to 100 Hz to match the sampling rate of the adult Sleep-EDF recordings, and then segmented into 30-second epochs with 15-second overlap. 

% For each subject, the resulting data were stored in a compressed NumPy \texttt{.npz} format with the following keys:

% \begin{itemize}
%     \item \texttt{x}: EEG data array of shape \texttt{(N, 3000, 2)} in $\mu$V units, where $N$ is the number of 30-second epochs and 2 represents the LEEG and REEG channels.
%     \item \texttt{y}: The corresponding sleep state labels were stored as a 1D array with shape \texttt{(N,)}, where each element is an integer representing the sleep stage: REM = 0, NREM = 1, and Intermediate = 2.
%     \item \texttt{fs}: Sampling frequency after downsampling (100 Hz).
%     \item \texttt{ch\_label}: String indicating the EEG channel configuration (e.g., \texttt{LEEG+REEG}).
%     \item \texttt{start\_datetime}: Datetime string representing the start time of the original EEG recording.
%     \item \texttt{file\_duration}: Total duration of the EEG recording in seconds.
%     \item \texttt{epoch\_duration}: Duration of each epoch used for the analysis (30 seconds).
%     \item \texttt{n\_all\_epochs}: Total number of 30-second epochs available in the processed recording.
%     \item \texttt{n\_epochs}: Number of epochs retained after label cleaning and quality control (should match \texttt{ n\_all\_epochs} if no segments were excluded).
% \end{itemize}

This unified preprocessing pipeline ensured that both handcrafted features and raw EEG signals were consistently prepared from the same ICA-cleaned and bandpass-filtered data. All downstream classification models—whether feature-based or end-to-end were trained and evaluated using the same annotated dataset.

For pretraining, we used the FetalSleepNet architecture but trained it on adult Sleep-EDF data to classify five stages (Awake, REM, N1, N2, and N3), excluding epochs labeled as MOVEMENT and UNKNOWN. This pretraining yielded CNN and long short-term memory (LSTM) weights that capture spatial and temporal EEG representations from adult sleep. During transfer to fetal data, we retained these pretrained weights and reinitialized only the final fully connected layer to output three fetal sleep states (REM, NREM, and Intermediate), enabling adaptation to the distinct label space of fetal sleep classification.

\subsection{Model Architecture}

\subsubsection{Rationale for Backbone Selection}
We adopt TinySleepNet~\cite{supratak2020tinysleepnet} as the backbone for our study due to its lightweight design (1.3M parameters) and strong performance in adult sleep staging, which reduces the risk of overfitting under data scarcity and aligns with the long-term goal of real-time, low-power monitoring. For context, existing sleep staging models trained on Sleep-EDF vary substantially in size: DeepSleepNet (21–22.9M)~\cite{supratak2017deepsleepnet}, SleepEEGNet (2.6M)~\cite{mousavi2019sleepeegnet}, XSleepNet (5.8M)~\cite{phan2021xsleepnet}, and SleepTransformer (3.8–3.9M)~\cite{phan2022sleeptransformer}. Among these, TinySleepNet is the most compact while retaining competitive accuracy.  

\subsubsection{Adaptation to Fetal EEG}
To adapt this architecture to fetal EEG, we developed FetalSleepNet, which introduces two key modifications:  
1) \textit{Dual-channel input:} designed to process bilateral parietal EEG (LEEG and REEG) from fetal sheep, corresponding to adult Fpz–Cz and Pz–Oz derivations.  
2) \textit{Enhanced temporal modeling:} the LSTM module was deepened from a single layer (128 units) to two stacked layers (256 units each) to better capture complex transitions, particularly for the ambiguous Intermediate state.  

\subsubsection{FetalSleepNet Architecture}
FetalSleepNet processes a sequence of dual-channel fetal EEG epochs and produces a sequence of sleep stage classifications of the same length in the many-to-many scheme. Formally, suppose there are $N$ EEG epochs $\{x_1, \ldots, x_N\}$ from dual-channel EEG, where $x_i \in \mathbb{R}^{E_s \times F_s \times 2}$, $E_s$ is the epoch duration (30 seconds) and $F_s$ is the sampling rate (100 Hz). Our model $f_\theta$ determines sleep stages for all epochs, resulting in $N$ predicted sleep stages $\{\hat{y}_1, \ldots, \hat{y}_N\}$, where $\hat{y}_i \in \{0,1,2\}$ corresponds to the three fetal sleep states: REM (0), NREM (1), and Intermediate (2). 

\begin{figure}[htbp]
    \centering
    \includegraphics[width=\linewidth]{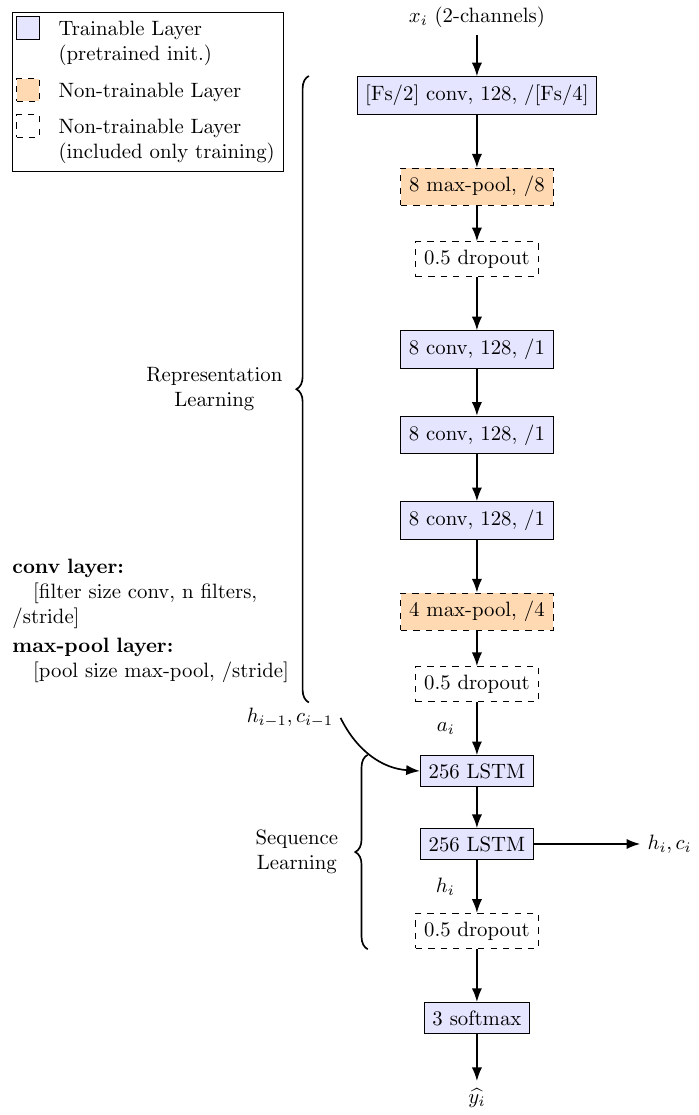}
    \caption{An overview architecture of FetalSleepNet. Each rectangular box
represents one layer in the model, and the arrows indicate the flow of data
from raw dual-channel EEG epochs ($x_i$) to sleep stages ($\hat{y}_i$).}
    \label{fig:fetalsleepnet}
\end{figure}

The model consists of two main parts: a convolutional network for representation learning and a recurrent network for sequence learning. The overall architecture is illustrated in Fig.~\ref{fig:fetalsleepnet}.

\paragraph{Representation Learning}
The first part of the network is a convolutional feature extractor. The CNN module consists of four convolutional layers, interleaved with two max-pooling and two dropout layers. The first convolutional layer uses a filter size of $F_s/2$ with stride $F_s/4$, followed by three convolutional layers with a filter size of 8 and stride 1. This block is designed to extract time-invariant features from raw dual-channel EEG epochs. Formally, the feature representation $a_i$ of the $i$-th input epoch $x_i$ is obtained as:
\begin{equation}
a_i = \mathrm{CNN}_{\theta_r}(x_i),
\end{equation}
where $\mathrm{CNN}_{\theta_r}$ denotes the convolutional layers parameterized by $\theta_r$.

\paragraph{Sequence Learning}
The second part of the network captures temporal dependencies across consecutive epochs. FetalSleepNet employs two stacked unidirectional LSTM layers, each with 256 hidden units, followed by a dropout layer. The stacked architecture increases the model’s capacity to capture complex temporal transitions, particularly those associated with the Intermediate state. Formally, the stacked LSTM is defined as:
\begin{align}
h_i^{(1)}, c_i^{(1)} &= \mathrm{LSTM}_{\theta_s^{(1)}}\left(h_{i-1}^{(1)}, c_{i-1}^{(1)}, a_i\right), \\
h_i^{(2)}, c_i^{(2)} &= \mathrm{LSTM}_{\theta_s^{(2)}}\left(h_{i-1}^{(2)}, c_{i-1}^{(2)}, h_i^{(1)}\right),
\end{align}
where $(h_i^{(1)}, c_i^{(1)})$ and $(h_i^{(2)}, c_i^{(2)})$ are the hidden and cell states of the first and second LSTM layers, respectively. The final hidden state $h_i^{(2)}$ is used as the input to the classification layer.

\paragraph{Output Layer}
The sequence learning module is followed by a fully connected (FC) softmax layer, which outputs the probability distribution over the three fetal sleep states: NREM, REM, and Intermediate.

\section{EXPERIMENTS}

\subsection{Overview}
To leverage knowledge from adult sleep data, we pretrained our FetalSleepNet architecture on the Sleep Cassette subset of the Sleep-EDF Expanded dataset. Ten-fold cross-validation was conducted on the adult data, and the best-performing checkpoint was selected as the pretrained model. During this stage, the output layer was configured for five adult sleep stages (Wake, REM, N1, N2, N3).

For transfer to fetal EEG, the final classification layer was reinitialized to classify three fetal sleep states (REM, NREM, and Intermediate), while the CNN and LSTM layers were initialized from the pretrained weights. We then evaluated three progressive transfer strategies:

\subsection{Baselines: CNN + LSTM on Fetal Data}
We first trained two CNN + LSTM models directly on the fEEG dataset:

\begin{itemize}
    \item \textbf{CNN + LSTM (raw EEG)}: The input consists of two synchronized EEG channels of filtered raw EEG signals. The model applies two 1D-CNN layers (16 and 32 filters, kernel size = 3, padding = 1) with batch normalization and ReLU activations, followed by max-pooling to halve the temporal resolution. The resulting feature sequence (feature dimension = 32) is fed into a unidirectional LSTM (hidden size = 64, 1 layer). The last time step output is passed through two fully connected layers (128 units and \textit{num\_classes} output). 
    \item \textbf{CNN + LSTM (handcrafted features)}: The input is a one-dimensional sequence of handcrafted EEG features
    %, including time-domain, frequency-domain, Hjorth, nonlinear, and coherence metrics
    . The architecture is identical to the raw EEG model except that the first convolutional layer has one input channel.
    \item \textbf{FetalSleepNet (fetal-only)}: 
    All layers are trained from scratch on fEEG without any external pretraining.
\end{itemize}

For all architectures, the ReLU activation function is applied after each convolutional and fully connected layer. These models serve as fetal-only baselines without any external data.

\subsection{FetalSleepNet: Transfer Learning from Adult Sleep EEG}
To leverage knowledge from adult sleep data, we pretrained our FetalSleepNet architecture on the Sleep Cassette subset of the Sleep-EDF Expanded dataset. Ten-fold cross-validation was conducted on the adult data, and the best-performing checkpoint was selected as the pretrained model.

We then transferred this model to fEEG via three progressive transfer strategies:

\begin{itemize}
    \item \textbf{Frozen CNN}: All convolutional layers were frozen; only the LSTM and FC layers were fine-tuned on fEEG.
    \item \textbf{Partial CNN fine-tuning}: The first CNN layer was unfrozen for adaptation, while the remaining CNN layers were kept frozen; LSTM and FC layers were fine-tuned.
    \item \textbf{Full CNN fine-tuning}: All CNN layers, along with the LSTM and FC layers, were fine-tuned on fEEG data.
\end{itemize}

\subsection{FetalSleepNet with Spectral Equalisation of Adult EEG}

To reduce the domain gap between adult and fEEG, we applied two preprocessing strategies to transform adult EEG into a fetal-compatible format:

\begin{itemize}
    \item First, we applied the same 1--22~Hz bandpass filter to adult EEG as used for fetal data~\cite{kelly2023progressive}.
    \item Next, we implemented a PSD-based spectral equalisation method to match the frequency characteristics of adult EEG to those of fEEG. 
    Spectral equalisation is a data-adaptive signal processing technique that applies frequency-dependent gain to correct relative spectral imbalances, 
    aiming to flatten or match a target spectrum without remapping frequencies~\cite{van1984spectral}. 
    In our case, the adult EEG power spectrum was scaled to the mean fetal spectrum, thereby reducing cross-domain spectral mismatch.
\end{itemize}

\subsubsection{Segmentation and Welch PSD}
% We first segmented each recording into contiguous, non-overlapping 30-s epochs.
% For PSD estimation in epoch $m$ of group $i\in\{\text{adult},\text{fetal}\}$ and channel $c$, we used Welch’s method~\cite{welch2003use} (sampling rate $f_s=100$ Hz; FFT length $N_p=512$; Hann window; 50\% overlap).
% Let $N_e=30f_s$ be the number of samples per epoch. Each epoch was partitioned into $R$ overlapping segments of length $N_p$ with hop $N_p/2$, i.e., $n_r=r(N_p/2)$ for $r=0,\ldots,R-1$, where
% \[
% R=\left\lfloor\frac{N_e-N_p}{N_p/2}\right\rfloor+1.
% \]
% (With $f_s=100$ Hz and $N_p=512$, $N_e=3000$ and $R=10$.) Define the windowed DFT of segment $r$ as
% \begin{equation}
% X^{(r)}_{i,m,c}[k]
% =\sum_{n=0}^{N_p-1} w[n]\,x_{i,m,c}[n_r+n]\,
% \mathrm{e}^{-\mathrm{j}2\pi kn/N_p}.
% \label{eq:segDFT}
% \end{equation}
% Here $w[n]$ is the Hann window, $U=\tfrac{1}{N_p}\sum_{n=0}^{N_p-1}w^2[n]$, and $f_k=\tfrac{k}{N_p}f_s$ for $k=0,\dots,N_p/2$ denotes one-sided frequency bins for real signals (one-sided scaling is applied: $\times2$ for $k=1,\dots,N_p/2-1$; DC and Nyquist not doubled).
% The Welch PSD for epoch $m$ is
% \begin{equation}
% \widehat{S}_{i,m,c}(f_k)
% =\frac{1}{R\,U\,N_p\,f_s}\sum_{r=0}^{R-1}\bigl\lvert X^{(r)}_{i,m,c}[k]\bigr\rvert^{2}.
% \label{eq:welch}
% \end{equation}
% Averaging across epochs yields the mean PSD
% \begin{equation}
% \mu_{i,c}(f_k)\approx \frac{1}{M_{i,c}}\sum_{m=1}^{M_{i,c}}\widehat{S}_{i,m,c}(f_k).
% \label{eq:mu}
% \end{equation}

Each recording was segmented into contiguous, non-overlapping 30-s epochs. 
For each epoch, the PSD was estimated using Welch’s method~\cite{welch2003use} 
with sampling rate $f_s=100$ Hz, FFT length $N_p=512$, Hann window, and 50\% overlap. 
The mean PSD across all epochs of group $i$ and channel $c$ was then computed as
\begin{equation}
\widehat{S}_{i,c}(f_k)\approx \frac{1}{M_{i,c}}
\sum_{m=1}^{M_{i,c}} \widehat{S}_{i,m,c}(f_k),
\label{eq:meanPSD}
\end{equation}
where $\widehat{S}_{i,m,c}(f_k)$ denotes the Welch PSD estimate for epoch $m$, 
$M_{i,c}$ is the number of available epochs for group $i$ and channel $c$, 
$i \in \{\text{adult}, \text{fetal}\}$ indexes the dataset group, 
$c$ indexes the EEG channel, and $f_k = kf_s/N_p$ ($k=0,\ldots,N_p/2$) denotes the frequency bin.

\subsubsection{Channel-wise Spectral Equalisation}
Let fetal EEG (fEEG) have left/right channels $\{\text{LEEG},\text{REEG}\}$ and adult EEG have midline derivations $\{\text{Fpz--Cz},\text{Pz--Oz}\}$. 
We map adult channels to fetal sides as
\[
\mathrm{map}(\text{Fpz--Cz})=\text{LEEG},\qquad \mathrm{map}(\text{Pz--Oz})=\text{REEG}.
\]
The frequency-dependent scaling factor is then defined as
\begin{equation}
s_c(f_k)=\frac{\widehat{S}_{\text{fetal},\,\mathrm{map}(c)}(f_k)}{\widehat{S}_{\text{adult},\,c}(f_k)+\epsilon},
\label{eq:sc}
\end{equation}
where $\widehat{S}_{i,c}(f_k)$ denotes the mean PSD at frequency $f_k$, 
$\epsilon=10^{-8}$ prevents division by zero, and $c\in\{\text{Fpz--Cz},\text{Pz--Oz}\}$. 
Let the adult spectrum be decomposed as
\begin{equation}
X_{\text{adult},\,c}[k]=A_c[k]\,\mathrm{e}^{\mathrm{j}\phi_c[k]},
\label{eq:adecomp}
\end{equation}
then applying amplitude-only spectral equalisation yields the fetal-style spectrum
\begin{equation}
\tilde{X}_{c}[k]=\sqrt{s_c(f_k)}\,X_{\text{adult},\,c}[k].
\label{eq:xtilde}
\end{equation}

% This approach assumes that the primary difference between adult EEG and fEEG lies in the relative distribution of spectral power across frequency bands, while the overall frequency structure and functional interpretation of canonical EEG rhythms (e.g., delta, theta) remains comparable across developmental stages. Although this is a simplifying assumption, it enables a lightweight and interpretable transformation to reduce domain shift. %Future work may investigate more flexible or non-linear alignment strategies if stronger spectral structural differences are identified.

% \subsubsection{Hermitian Symmetry and IFFT}
% To guarantee real-valued outputs, we enforce Hermitian symmetry before the IFFT:
% \begin{subequations}\label{eq:hermitian}
% \begin{align}
% \Im\{\tilde{X}_{c}[0]\}&=\Im\{\tilde{X}_{c}[N_p/2]\}=0, \label{eq:hermitian:a}\\
% \tilde{X}_{c}[N_p-k]&=\tilde{X}_{c}[k]^{\ast},\quad k=1,\ldots,N_p/2-1. \label{eq:hermitian:b}
% \end{align}
% \end{subequations}
% The $N_p$-point IFFT then yields the time-domain segment
% \begin{equation}
% \tilde{x}_{c}[n]=\frac{1}{N_p}\sum_{k=0}^{N_p-1}\tilde{X}_{c}[k]\,
% \mathrm{e}^{\mathrm{j}2\pi kn/N_p},\quad n=0,\ldots,N_p-1.
% \label{eq:ifft}
% \end{equation}

\subsubsection{Hermitian Symmetry and IFFT}
To guarantee real-valued outputs, we enforce Hermitian symmetry before the IFFT:
\begin{subequations}\label{eq:hermitian}
\begin{align}
\Im\{\tilde{X}_{c}[0]\}&=\Im\{\tilde{X}_{c}[N_p/2]\}=0, \label{eq:hermitian:a}\\
\tilde{X}_{c}[N_p-k]&=\tilde{X}_{c}[k]^{\ast},\quad k=1,\ldots,N_p/2-1. \label{eq:hermitian:b}
\end{align}
\end{subequations}
Here the scaling factor $s_c(f_k)$ (Eq.~\ref{eq:sc}) is applied directly in the frequency domain, 
and Hermitian symmetry is used to reconstruct the negative-frequency bins, ensuring that 
the inverse FFT produces a real-valued time-domain signal:
\begin{equation}
\tilde{x}_{c}[n]=\frac{1}{N_p}\sum_{k=0}^{N_p-1}\tilde{X}_{c}[k]\,
\mathrm{e}^{\mathrm{j}2\pi kn/N_p},\quad n=0,\ldots,N_p-1.
\label{eq:ifft}
\end{equation}

Although both PSD-informed spectral equalisation and bandpass filtering are linear operations and theoretically order-invariant, we applied spectral equalisation before filtering to ensure that the equalisation was applied over the full frequency spectrum of the original signal. This order avoids prematurely removing frequency components that may contribute to the spectral distribution, particularly near the filter cutoffs. Importantly, the transformation was performed on the entire continuous EEG signal rather than on segmented epochs, preserving temporal continuity and avoiding boundary artifacts. Bandpass filtering was applied only after spectral equalisation to maintain consistency with the fEEG preprocessing pipeline.

\begin{figure}[htbp]
    \centering
    \subfloat[Fetal LEEG vs REEG]{
        \includegraphics[width=\linewidth]{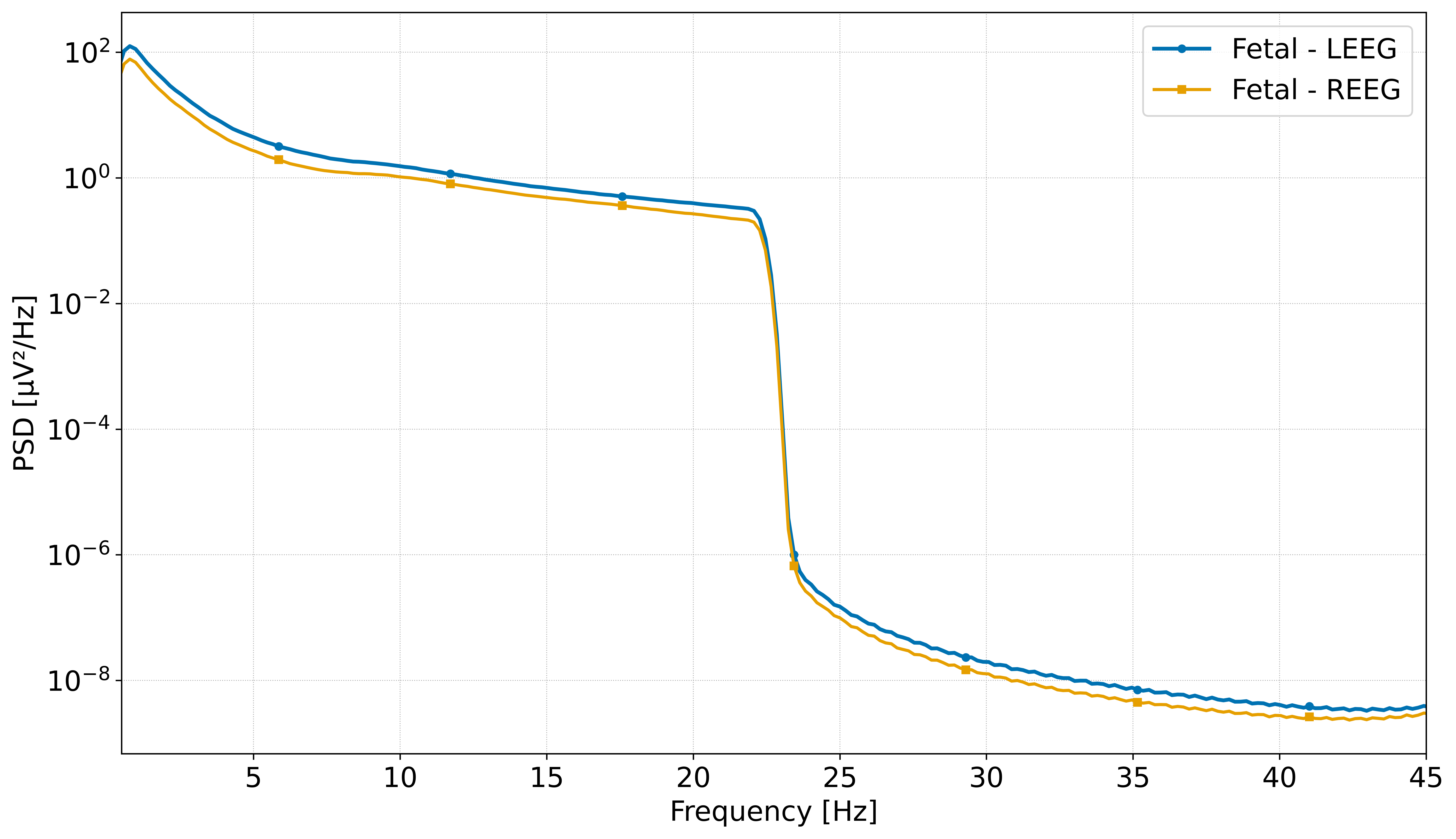}
        \label{fig:PSD_Fetal_LEEG_vs_REEG_2lines}
    }
    \vspace{0.4cm}
    \subfloat[Adult raw vs equalised]{
        \includegraphics[width=\linewidth]{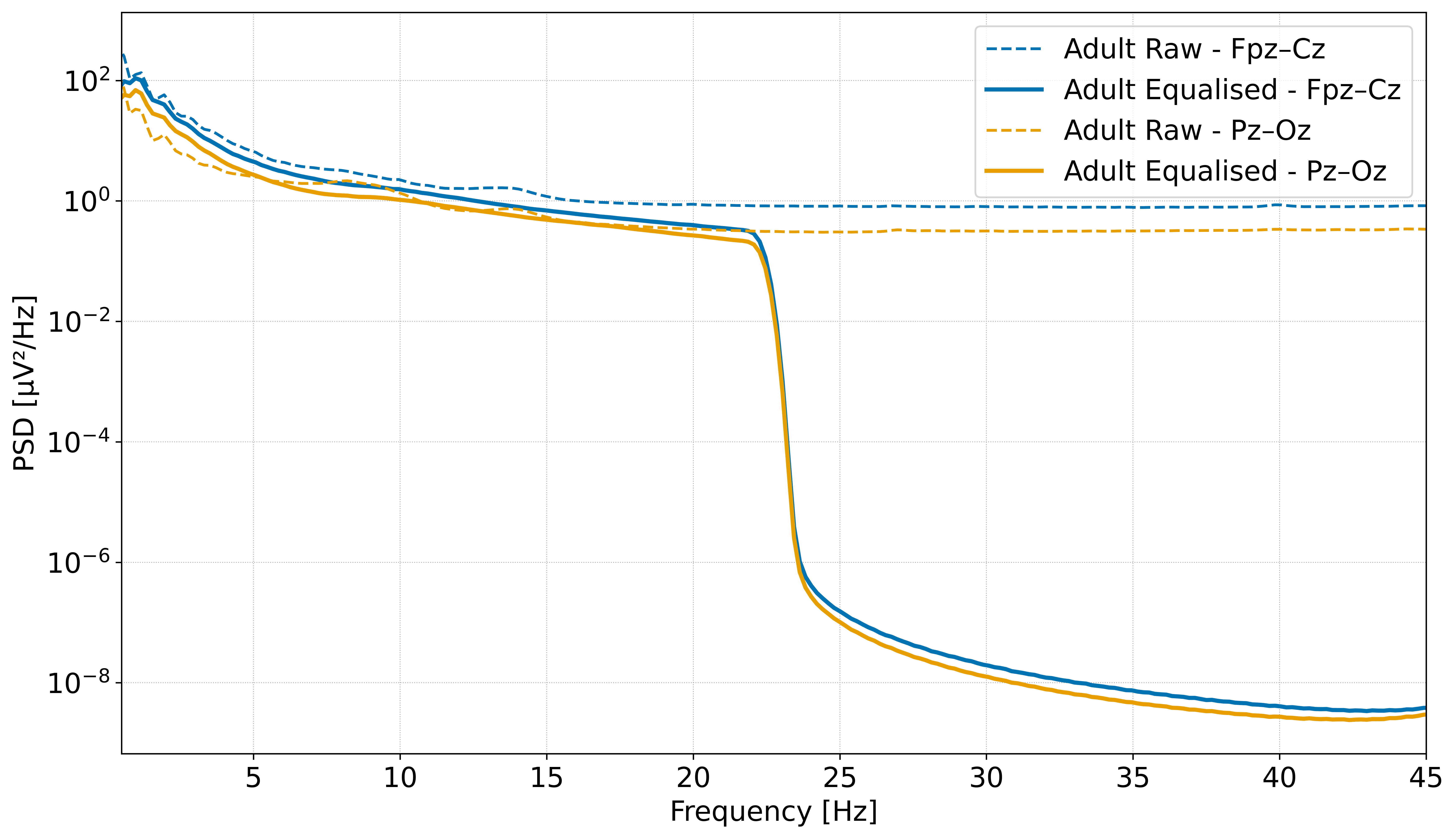}
        \label{fig:PSD_Adult_Raw_vs_Scaled_4lines}
    }
    \caption{Comparison of PSD profiles for fetal and adult EEG. 
    (a) fEEG PSD for bilateral channels: LEEG (blue, solid with circles) and REEG (orange, dashed with squares) show closely matched spectral profiles across 0.5–45 Hz, indicating strong left–right consistency. 
    (b) Adult EEG PSD before and after PSD-based spectral equalisation without filtering: Fpz–Cz and Pz–Oz are shown in blue and orange, respectively; raw = dashed, scaled = solid. After equalisation, the adult spectra closely resemble the fetal spectra across the frequency range of interest.}
    \label{fig:PSD_Fetal_vs_Adult}
\end{figure}

The fetal spectra from the two cortical channels are highly consistent. As shown in Fig.~\ref{fig:PSD_Fetal_LEEG_vs_REEG_2lines}, LEEG and REEG exhibit very similar power–frequency profiles across the band of interest, with only small lateral differences. This stable bilateral pattern provides a reliable target distribution for cross-domain adaptation.

Building on this observation, we applied PSD-based spectral equalisation to full-spectrum adult EEG without bandpass filtering. After scaling, the adult Fpz–Cz and Pz–Oz spectra closely resemble the fetal LEEG/REEG profiles (Fig.~\ref{fig:PSD_Adult_Raw_vs_Scaled_4lines}), while preserving physiologically informative features (e.g., sleep spindles, K-complexes~\cite{rechtschaffen1968}).

\subsection{Experimental Setup}
% All experiments were conducted on a workstation running Ubuntu 18.04.6 LTS, equipped with an Intel® Core™ i9-9900K CPU @ 3.60GHz × 16, a single NVIDIA Titan RTX GPU (24GB memory), CUDA 10.0, and cuDNN 7. The model training was implemented using TensorFlow 1.13.1. Each model was trained using the Adam optimizer with an initial learning rate of $1 \times 10^{-3}$, $\beta_1 = 0.9$, $\beta_2 = 0.999$, and $\epsilon = 10^{-8}$. Gradient clipping was applied with a maximum norm of 5.0 to stabilize training. We used early stopping based on the macro F1 score with a patience of 30 evaluations.

\subsubsection{Hardware and Software}
All experiments were conducted on a workstation running Ubuntu 18.04.6 LTS, equipped with an Intel Core i9-9900K CPU @ 3.60GHz × 16, a single NVIDIA Titan RTX GPU (24GB memory), CUDA 10.0, cuDNN 7, and TensorFlow 1.13.1.

\subsubsection{Hyperparameter Tuning}
Hyperparameters (initial learning rate, $\beta$ values, gradient clipping threshold, and early stopping patience) were selected via grid search on the validation set in the first leave-one-subject-out (LOSO) fold, and the resulting configuration was applied to all folds to avoid test set leakage. For the learning rate, we observed that overly large values led to rapid convergence but risked missing the optimal point, while overly small values slowed training; a value of $1 \times 10^{-3}$ provided the best trade-off. For early stopping patience, we tested multiple settings, finding that excessively large patience tended to cause overfitting, whereas too small patience led to underfitting; a value of 30 evaluations achieved the most stable performance.

\subsubsection{Architecture Details}
The batch size was set to 256, and each model was trained for up to 200 epochs. For the CNN + LSTM architecture, we used two bidirectional LSTM layers with 256 hidden units. In transfer learning experiments, we initialized the model with pretrained weights obtained from adult sleep data. To avoid overfitting, no signal augmentation was applied to fEEG data.

\subsubsection{Weighted Cross-Entropy Loss for Class Imbalance}
To address class imbalance, we employed a weighted cross-entropy loss:
\begin{equation}
\mathcal{L}_{\text{WCE}} =
\frac{\sum_{i=1}^{N} w^{\text{seq}}_{i} \cdot w^{\text{cls}}_{y_i} \cdot \ell_i}
{\sum_{i=1}^{N} w^{\text{seq}}_{i}}, \quad
\ell_i = -\log p_{i,y_i}.
\end{equation}
Here, $p_{i,y_i}$ is the predicted probability for the true label $y_i$.
The class weight $w^{\text{cls}}_{y_i}$ was computed from the class
distribution of the training split in each fold using the balanced scheme
\begin{equation}
w^{\text{cls}}_{c} = \frac{N_{\text{tr}}}{C \cdot n_c}, \quad 
c \in \{\text{NREM}, \text{REM}, \text{Intermediate}\},
\end{equation}
where $N_{\text{tr}}$ is the total number of training samples in the
current fold, $C=3$ is the number of classes, and $n_c$ is the number
of samples belonging to class $c$. This formulation automatically
assigns larger weights to minority classes (e.g., Intermediate) and
smaller weights to majority classes (e.g., REM), while keeping the
average weight close to one. Sequence weights $w^{\text{seq}}_{i}$
were introduced in our formulation to allow equal contribution from
subjects with different recording lengths; however, they were not
applied in the final experiments and were set to $w^{\text{seq}}_{i}=1$
for all samples. Unlike oversampling or class-balanced pretraining,
this strategy preserves the natural temporal structure of sleep
recordings while mitigating the impact of imbalance at the loss level.

\subsection{Evaluation}
Model performance was evaluated using LOSO cross-validation, where each of the 24 subjects served once as the test set. In each fold, two subjects were used for validation and the remaining 21 for training. Predictions were made at the epoch level, with each 30-second EEG segment assigned a single sleep state label (NREM, REM, or Intermediate). All reported metrics were computed on these per-epoch predictions.

For each class $c \in \{\text{NREM}, \text{REM}, \text{Intermediate}\}$, the F1 score was computed, and the macro F1 was obtained.%:
%\begin{equation}
%\text{Macro F1} = \frac{1}{3} \sum_{c} F1_c
%\end{equation}
 It was chosen to account for class imbalance and provide a more robust estimate of overall performance.

\begin{table*}[htbp]
\caption{Fetal sleep state classification performance across models, input types, and training strategies. \\Values with $\pm$ indicate mean $\pm$ standard deviation (\%).}
\label{tab:classification_comparison}
\begin{adjustbox}{width=\textwidth}
\begin{tabular}{lcccccccc}
\hline
\textbf{Model/Setting} & \textbf{Pretraining Dataset} & \textbf{Input Type} & \textbf{Training Strategy} & \textbf{Accuracy (\%)} & \textbf{Macro F1} & \textbf{NREM F1} & \textbf{REM F1} & \textbf{Intermediate F1} \\
\hline
CNN + LSTM & None (Fetal Only) & Raw EEG & Baseline & 76.43 $\pm$ 10.58 & 54.89 $\pm$ 10.43 & 76.78 $\pm$ 14.8 & 77.32 $\pm$ 18.95& \textbf{10.56 $\pm$ 7.61} \\
CNN + LSTM & None (Fetal Only) & Handcrafted Features & Baseline & 82.6 $\pm$ 9.97 & 58.27 $\pm$ 6.64 & 82.3 $\pm$ 11.92 & 85.7 $\pm$ 10.09 & 6.8 $\pm$ 6.10 \\
FetalSleepNet & None (Fetal Only) & Raw EEG & Baseline & 80.1 $\pm$ 9.85 & 57.33 $\pm$ 10.12 & 81.5 $\pm$ 13.92 & 82.4 $\pm$ 14.05 & 8.1 $\pm$ 8.12\\
\hline
FetalSleepNet & Sleep-EDF (Adult) & Raw EEG & Transfer (Frozen CNN) & $\approx$ 18.7 & $\approx$ 18.7 & --- & --- & --- \\
FetalSleepNet & Sleep-EDF (Adult) & Raw EEG & Transfer (Partial CNN) & 72.8 $\pm$ 15.99 & 51.8 $\pm$ 15.46 & 79.6 $\pm$ 34.06 & 69.7 $\pm$ 13.38 & 6.1 $\pm$ 8.19 \\
FetalSleepNet & Sleep-EDF (Adult) & Raw EEG & Transfer (Full CNN) & 84.0 $\pm$ 10.35 & 61.43 $\pm$ 6.90 & 88.1 $\pm$ 10.52 & 87.0 $\pm$ 10.85 & 9.2 $\pm$ 9.80 \\
\hline
\textbf{FetalSleepNet} & \textbf{Sleep-EDF (Adult)} & \textbf{Spectral-Equalised EEG} & \textbf{Transfer (Full CNN)} & \textbf{86.6 $\pm$ 4.77} & \textbf{62.47 $\pm$ 6.04} & \textbf{88.1 $\pm$ 5.53} & \textbf{89.8 $\pm$ 6.43} & 9.5 $\pm$ 14.34 \\
\hline
\end{tabular}
\end{adjustbox}
\end{table*}

\section{RESULTS}

\subsection{Overall Performance Comparison}

Table~\ref{tab:classification_comparison} summarizes the performance of all models across input types, training strategies, and pretraining datasets (mean ± standard deviation across LOSO folds). Among fetal-only models trained from scratch, the CNN+LSTM using handcrafted features achieved the highest macro F1 score (58.27 ± 6.64\%), outperforming the same architecture trained on raw EEG (54.89 ± 10.43\%). The FetalSleepNet architecture, when trained solely on fetal raw EEG, achieved a macro F1 of 57.33 ± 10.12\%, showing slightly better balance than CNN+LSTM on raw EEG but still below the handcrafted-feature baseline.

Direct transfer from adult Sleep-EDF without adaptation (frozen CNN) resulted in very poor performance (approximately 18.7\%), confirming a substantial domain gap between adult EEG and fEEG. Gradually increasing the number of trainable CNN layers improved results, with full CNN fine-tuning on raw fEEG reaching 61.43 ± 6.90\% macro F1—surpassing all fetal-only baselines. This demonstrates that transfer learning can be highly beneficial when the pretrained representation is fully adapted to the target domain.

Applying spectral equalisation to adult EEG before transfer yielded the highest overall performance. The spectrally-equalised transfer model achieved 62.47 ± 6.04\% macro F1 and 86.6 ± 4.77\% accuracy, with class-wise F1 scores of 88.1 ± 5.53\% (NREM), 89.8 ± 6.43\% (REM), and 9.5 ± 14.34\% (Intermediate). While the handcrafted-feature CNN+LSTM still performed best on the Intermediate state (10.56 ± 7.61\%), the spectrally-equalised transfer model provided the best balance across all classes, highlighting the advantage of reducing spectral mismatch before transfer.

\subsection{Quantitative Comparison of Top Models With and Without Spectral Equalisation}

Among all tested configurations, two models achieved the highest overall performance:
\begin{enumerate}
    \item Model R: FetalSleepNet trained on Sleep-EDF (Adult) \emph{raw} EEG and transferred using the full CNN.
    \item Model S: FetalSleepNet trained on Sleep-EDF (Adult) \emph{spectral-equalised} EEG and transferred using the full CNN.
\end{enumerate}

Although absolute performance differences were modest, we conducted a paired statistical test to assess whether spectral equalisation (Model~S) yields a significant improvement over the raw-EEG transfer baseline (Model~R). This analysis quantifies the benefit of spectral equalisation when adapting adult sleep EEG to fetal sleep staging.

We employed the Wilcoxon signed-rank test~\cite{wilcoxon1945individual}, a non-parametric alternative to the paired $t$-test that does not assume normality of the paired differences (it does assume symmetry of the difference distribution). Pairs were defined at the \emph{fold} level under LOSO, yielding $n=24$ paired observations (one fold per subject) for each metric. 
%For clarity, let $x_i$ denote the metric of Model~S (spectral-equalised) and $y_i$ that of Model~R (raw) on the same fold $i$; thus a positive difference $d_i=x_i-y_i$ favors Model~S.

%Given two paired samples $\{x_i\}$ and $\{y_i\}$ ($i=1,\dots,24$), the test proceeds as follows:
%\begin{enumerate}
%     \item Compute paired differences $d_i=x_i-y_i$ and exclude pairs with $d_i=0$.
%     \item Rank the absolute differences $|d_i|$ from smallest to largest (average ranks for ties).
%     \item Assign the sign of $d_i$ to each rank and sum signed ranks separately to obtain
%     \begin{align}
%         W^+ &= \sum_{\{i: d_i>0\}} R_i, \\
%         W^- &= \sum_{\{i: d_i<0\}} R_i.
%     \end{align}
%     \item The test statistic is $W=\min(W^+,W^-)$.
% \end{enumerate}
%We tested the two-sided null hypothesis $H_0:\ \text{median}(d_i)=0$. Small $W$ together with a low two-sided $p$-value indicates a significant difference between the two models. 
Fold-wise observations are independent across folds by construction (each fold holds out a distinct subject).
%\subsubsection{Results}
We applied the test to fold-wise metrics from LOSO-CV ($n=24$ pairs) for Accuracy, Macro~F1, REM~F1, NREM~F1, and Intermediate~F1. Two-sided exact $p$-values are reported; conclusions remain unchanged after Holm--Bonferroni correction~\cite{holm1979simple} for five comparisons ($\alpha=0.05$).

\begin{table}[htbp]
\caption{Wilcoxon signed-rank test comparing Model~S (spectral-equalised) vs.\ Model~R (raw) on LOSO fold-wise metrics ($n=24$ pairs). Two-sided exact $p$-values are shown; ``Significant'' denotes $p<0.05$.}
\label{tab:wilcoxon_results}
\centering
\begin{tabular}{lccc}
\hline
\textbf{Metric} & \textbf{$W$ statistic} & \textbf{$p$-value} & \textbf{Significance} \\
\hline
Accuracy        & 0.0 & $<0.0001$ & \textbf{Significant} \\
Macro F1        & 0.0 & $<0.0001$ & \textbf{Significant} \\
REM F1          & 0.0 & $<0.0001$ & \textbf{Significant} \\
NREM F1         & 0.0 & $<0.0001$ & \textbf{Significant} \\
Intermediate F1 & 1.0 & $0.0002$  & \textbf{Significant} \\
\hline
\end{tabular}
\end{table}

%\subsubsection{Interpretation}
For Accuracy, Macro~F1, REM~F1, and NREM~F1, $W=0$ indicates that all non-zero paired differences favored Model~S across all folds, yielding highly significant results ($p<0.0001$). For Intermediate~F1, $W=1$ reflects a single fold with the opposite direction, yet the result remains statistically significant ($p=0.0002$). Collectively, these findings provide strong and consistent evidence that spectral equalisation confers a measurable and statistically significant advantage when transferring from adult sleep EEG to fetal sleep staging.

\subsection{Sleep Stage Distribution}
To better understand the class distribution differences between domains, we first examined the relative frequencies of sleep stages in the fetal and adult datasets.

Physiologically, adult N1–N3 correspond to fetal NREM, adult REM corresponds to fetal REM, and adult Wake (W) has no direct fetal analogue. The fetal Intermediate state represents a heterogeneous category, including short sleep cycles ($<$ 3 min), mismatched EEG amplitude and SEF frequency, and transitional EEG patterns showing mixed REM–NREM features, but it does not correspond to wakefulness.

In the fetal dataset, REM accounts for the largest proportion 
(51.5\%, 7,602 epochs), followed by NREM (40.5\%, 5,975) and 
Intermediate (8.0\%, 1,188). In contrast, adult NREM-like stages 
collectively dominate (52.0\%, 103,693)\footnote{N1 (10.8\%, 21,522) + N2 (34.7\%, 69,132) + N3 
(6.5\%, 13,039).}, followed by Wake 
(35.0\%, 69,824) and REM (13.0\%, 25,835). This marked mismatch in both label composition and class balance—particularly the much higher REM proportion in fetal data and the substantial Wake proportion in adult data—introduces a label-prior and spectral-content gap between domains. These disparities motivate the application of spectral equalisation and other domain adaptation techniques to reduce distributional shifts and improve cross-domain transfer performance.

% \begin{table}[htbp]
% \centering
% \caption{Sleep stage distribution}
% \label{tab:stage_distribution}
% \begin{tabular}{lcc}
% \hline
% \textbf{Stage} & \textbf{Fetal} & \textbf{Adult (\% , n)} \\
% \hline
% NREM           & 40.5\% (5,975)  & 52.0\% (103,693)$^\ast$ \\
% REM            & 51.5\% (7,602)  & 13.0\% (25,835) \\
% Intermediate   & 8.0\% (1,188)   & -- \\
% Wake           & --              & 35.0\% (69,824) \\
% \hline
% \end{tabular}
% \\[2pt]
% {\footnotesize $^\ast$Adult NREM = N1 (10.8\%, 21,522) + N2 (34.7\%, 69,132) + N3 (6.5\%, 13,039).}
% \end{table}

\subsection{Training Dynamics with Spectral Equalisation}

\begin{figure}[htbp]
    \centering
    \includegraphics[width=\linewidth]{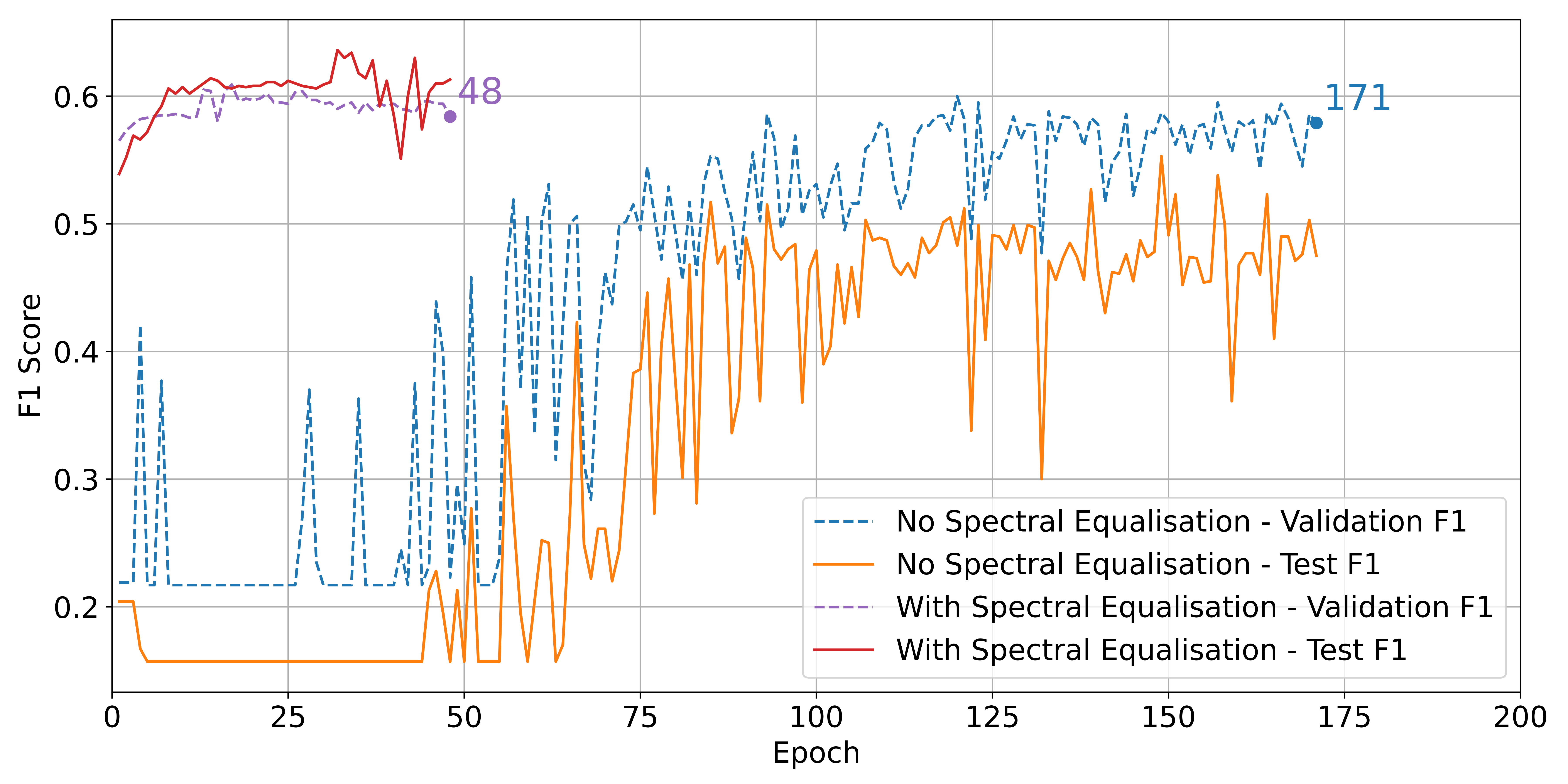}
    \caption{Comparison of validation and test F1 scores over training epochs for one fold. The model trained with spectral equalisation (purple and red) converged earlier (epoch 48) than the model trained without spectral equalisation (blue and orange, epoch 171), while achieving comparable or better performance.}
    \label{fig:21165_f1_comparison.png}
\end{figure}

Fig.~\ref{fig:21165_f1_comparison.png} illustrates the training trajectories of validation and test F1 scores for one fold, comparing models trained with and without spectral equalisation. Beyond the difference in convergence speed—epoch 48 versus epoch 171—the figure also highlights two additional aspects: (i) the spectral equalisation model (purple and red lines) exhibits a smoother and more stable trajectory across both validation and test sets, whereas the model without spectral equalisation (blue and orange lines) fluctuates substantially before converging; and (ii) spectral equalisation consistently maintains higher test F1 throughout training, suggesting improved generalization. These dynamics, visible only through the training curves, demonstrate that spectral equalisation not only accelerates convergence but also stabilizes learning and improves robustness.

% Despite the shorter training duration, the spectrally aligned model achieved comparable or better performance on the validation set and demonstrated more stable test F1 scores across epochs. This indicates that aligning the spectral distribution between adult EEG and fEEG can accelerate convergence and improve generalization, likely by reducing domain shift in the frequency domain. These findings support the effectiveness of spectral equalisation as a lightweight yet impactful preprocessing step in cross-domain sleep stage classification.

\subsection{Computation Time}
We benchmarked the inference latency of FetalSleepNet on both CPU and GPU using a system equipped with an Intel i9-9900K CPU (3.60\,GHz, 16 threads) and an NVIDIA TITAN RTX GPU. 
Table~\ref{tab:comp_time} reports the latency statistics averaged over all folds of our LOSO evaluation across 24 fetal sheep.

\begin{table}[h]
\centering
\caption{Computation time comparison of FetalSleepNet.}
\label{tab:comp_time}
\begin{tabular}{lcccc}
\hline
Device & Avg & Throughput & Min & Max \\
       & (ms/epoch) & (epochs/s) & (ms) & (ms) \\
\hline
CPU (i9-9900K) & 5.666 & 176.5 & 3.331 & 38.171 \\
GPU (TITAN RTX) & 6.503 & 153.8 & 2.126 & 747.648 \\
\hline
\end{tabular}
\end{table}

As shown, the CPU achieved a slightly lower average per-epoch latency than the GPU due to the small model size and batch size (set to one). Here, a 30-second epoch corresponds to one input segment used for a single sleep state classification. Nevertheless, both devices demonstrated real-time performance ($<$ 10 ms per 30-second epoch), confirming the efficiency of the proposed framework. These results further highlight FetalSleepNet’s lightweight design, making it well suited for deployment in low-power, real-time, and wearable fetal monitoring systems.

\section{DISCUSSION}
In this study, we aimed to develop the first automated framework for classifying fetal sheep sleep stages using EEG. By leveraging transfer learning, we adapted a compact adult sleep classifier to the fetal domain. Our results demonstrate that automated fetal sleep staging is feasible, achieving strong performance in distinguishing the principal sleep states of late-gestation fetal sheep. This is an important step forward, as reliable automated methods have the potential to accelerate fetal neurodevelopmental research by replacing time-consuming manual scoring. To our knowledge, this is the first model in the literature to enable automated classification of fetal sheep sleep, highlighting both the feasibility and the translational significance of cross-domain approaches for fetal monitoring.

\subsection{Performance on Intermediate State}
The Intermediate sleep state was the most difficult to classify across all models. 
The fetal-only baseline achieved the highest Intermediate F1 (10.56\%) when using CNN + LSTM on raw EEG. 
However, when applying transfer learning with spectral equalisation, Intermediate F1 reached 9.5\%—closely approaching this fetal-only baseline peak, while achieving far better overall accuracy and REM/NREM performance. 

\begin{figure}[t]
    \centering
    \includegraphics[width=0.9\linewidth]{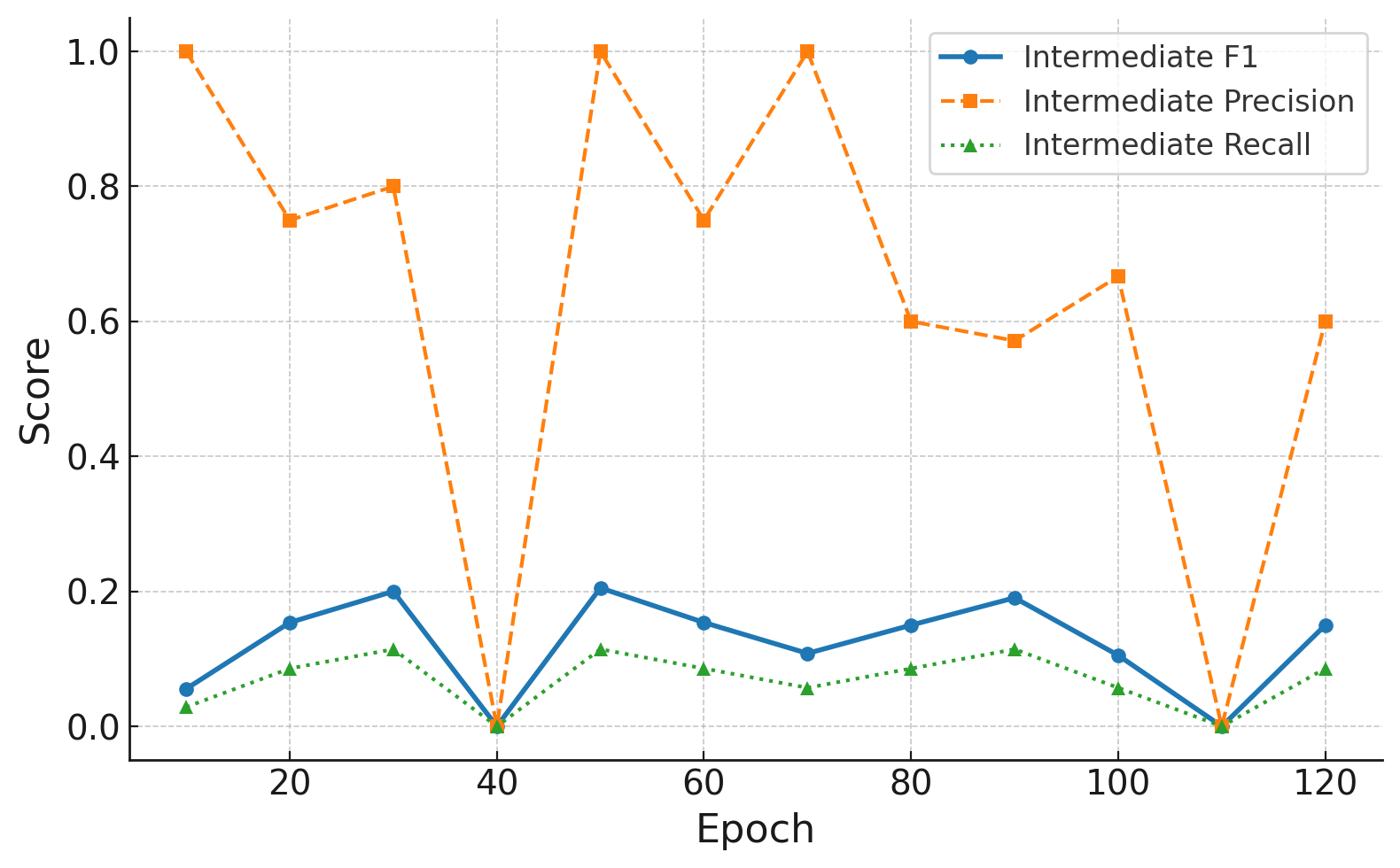}
    \caption{Per-epoch performance of the Intermediate sleep state across 120 epochs.
    Metrics were computed every 10 epochs using confusion matrices from the
    transfer learning setting with spectral equalisation (Full CNN). 
    The curves show F1 score (solid), precision (dashed), and recall (dotted). 
    Performance on the Intermediate state remains low and stable, reflecting its
    transient nature and overlapping features with REM and NREM.}
    \label{fig:intermediate_epoch_curve}
\end{figure}

This indicates that while transfer learning improves general performance, distinguishing Intermediate states remains a persistent challenge. 
The difficulty is not attributable to insufficient training, but rather to the nature of this class: The Intermediate state encompasses transitional periods between NREM and REM, as well as other ambiguous epochs such as short sleep cycles ($<$ 3 min) or mismatched EEG amplitude and frequency. It is relatively infrequent and typically characterized by overlapping EEG features.
Our per-epoch analysis (Fig.~\ref{fig:intermediate_epoch_curve}) shows that performance on the Intermediate state remains consistently poor across training. While F1 scores fluctuate around a low baseline, neither precision nor recall exhibits a clear upward trajectory, reflecting the difficulty of learning this class. Rather than indicating a lack of training, these unstable patterns highlight the intrinsic ambiguity and scarcity of Intermediate epochs, which limit the model’s ability to converge to a robust decision boundary. Thus, spectral equalisation does not specifically enhance Intermediate classification, but it enables strong overall generalization, yielding robust performance on REM and NREM states despite the intrinsic ambiguity of the Intermediate state.

\subsection{Impact of Spectral Equalisation on Training Dynamics}
The observed acceleration in convergence and the more stable trajectories of the spectrally equalised model suggest that spectral equalisation mitigates cross-domain discrepancies between adult EEG and fEEG. Rather than shifting or warping frequencies, spectral equalisation re-weights the spectral content of adult EEG to match the distribution of fEEG, thereby reducing domain shift in the frequency domain. This adjustment provides the model with inputs that are more consistent across domains, which facilitates transfer learning. From a practical standpoint, spectral equalisation not only shortens training time but also improves generalization to unseen fetal data, addressing one of the major challenges in data-scarce settings. Importantly, its lightweight nature makes spectral equalisation a scalable and low-cost component that can be readily integrated into future cross-domain EEG transfer pipelines.

\subsection{Interpretability of Handcrafted Features}
The superior performance of the handcrafted-feature baseline under limited data conditions suggests that physiologically meaningful features, distilled into a compact representation, are more readily learnable by the model than raw EEG. Such features led to more accurate classification of both REM and NREM states, whereas end-to-end models trained directly on raw EEG must learn these representations from the signal itself, typically requiring substantially more training data to achieve comparable performance.

To further probe the handcrafted-feature baseline, we assessed feature contributions using permutation importance, defined as the macro-F1 drop after shuffling a single feature on the test set (5 repeats; averaged across LOSO folds). The top five contributors were: (1) PFD of the R~EEG channel; (2) relative $\alpha$-band (8–13\,Hz) power of the R~EEG channel; (3) PFD of the L~EEG channel; (4) relative $\beta$-band (13–22\,Hz) power of the R~EEG channel; and (5) DFA of the L~EEG channel. These results indicate that nonlinear signal complexity (PFD), band-limited spectral content ($\alpha/\beta$), and long-range temporal structure (DFA) carry most of the discriminative information for fetal sleep staging, with a mild right-hemisphere emphasis.

\subsection{Practical Considerations for Real-World Deployment}
In real-world deployment, per-subject, per-channel z-score normalization can be achieved via a one-time initial calibration rather than continuous offline computation. The intrinsic sleep cycle of fetal sheep lasts approximately 20–40 minutes with an approximately 1:1 NREM:REM ratio~\cite{ruckebusch1977sleep,szeto1985prenatal}. In our fetal sheep dataset, cycle durations ranged from about 10 to 40 minutes, with the majority falling within the 20–40 minute window. We therefore recommend an initial calibration period of up to 40 minutes, corresponding to the maximum cycle length, to ensure both major sleep states are sampled and representative normalization parameters are obtained. After this stage, the system can operate in near-real-time, keeping the normalization parameters fixed or updating them slowly via an exponential moving average with a long time constant (on the order of hours) to track gradual drift without introducing state-dependent bias. When sufficient calibration is not feasible, population/device priors, robust scale estimators (e.g., median/MAD), and artifact rejection can be applied to maintain stability.

In addition to normalization strategies that enable stable near-real-time operation, the computational design of the model is also critical for deployment. TinySleepNet was originally designed as a compact and efficient architecture, and FetalSleepNet retains this property. Its small parameter count and low inference latency ensure feasibility for continuous monitoring on standard hardware, supporting scalability in large experimental studies and potential integration into clinical pipelines.

\subsection{Translation to Human Applications}

Late-gestation fetal sheep exhibit sleep state organization, EEG spectral features, and state transition dynamics that are broadly analogous to those of human fetuses at comparable developmental stages~\cite{nijhuis1982there,szeto1985prenatal,tang2025fetalsleepcrossspeciesreview}. This similarity underpins the value of fetal sheep sleep classification as a translational model, providing a brain-based reference for evaluating autonomic and neural development in utero. In humans, direct fEEG acquisition is not currently feasible, due to the invasive approach required for accurate signal acquisition, and is typically only possible during labor using scalp electrodes in exceptional clinical or research contexts~\cite{thaler2000real}. Such recordings are therefore extremely scarce and impractical for continuous, non-invasive monitoring or large-scale automated detection systems. Consequently, most human studies rely on indirect measures such as cardiotocography, fECG or fMCG to infer neural or sleep states~\cite{tang2025fetalsleepcrossspeciesreview}, making the high-quality direct fEEG data from fetal sheep an indispensable surrogate for developing and validating brain-based fetal monitoring approaches.

By establishing automated and objective EEG-based sleep state classification in fetal sheep, our FetalSleepNet provides a robust “gold standard” for evaluating how HRV measures vary across states and developmental stages. These brain-based benchmarks can guide the adaptation of non-invasive modalities (e.g., Doppler ultrasound, fECG) toward human applications, supporting future translational studies.
%Among existing proxies, fECG is the most clinically practical but its signal quality is reduced at 28–34 weeks due to vernix caseosa~\cite{wallwitz2012development}. In contrast, 
In particular, fMEG provides a non-invasive, brain-based modality that can serve as a bridge for transfer learning from sheep and neonatal EEG to human fetal EEG, 
%while fMCG offers a practical cardiac surrogate 
if cost and accessibility challenges are addressed. Translation to clinical practice will therefore require robust artifact suppression, adaptation to non-invasive modalities, and validation in human fetal or preterm cohorts. Behavioral state classification in human studies is also less direct than in animal models, often relying on heart rate patterns rather than multimodal observations~\cite{wallwitz2012development}.

Beyond technical feasibility, the ultimate clinical goal is to improve risk prediction of fetal and maternal complications by integrating state-aware analyses across large cohorts. Integrating HRV metrics with established indices such as the fetal autonomic brain age score \cite{hoyer2014fetal} could enable early detection of conditions like fetal growth restriction, hypoxia or congenital infections. Large-scale, longitudinal validation—supported by our FetalSleepNet as a robust benchmark for brain-based state classification—will be essential to establish normative developmental trajectories and clinically meaningful thresholds \cite{schneider2018developmental}.

\subsection{Limitations}
Our spectral equalisation method assumes that the primary difference between adult EEG and fEEG lies in the relative distribution of spectral power, while canonical rhythms (e.g., delta/theta activity) remain broadly comparable. Evidence from fetal sheep studies supports the presence of adult-like electrocortical rhythms by approximately 120 days of gestation~\cite{ruckebusch1977sleep,szeto1985prenatal}, but also highlights developmental differences in the precise frequency boundaries, the relative prevalence of REM sleep, and frequent arousal$\rightarrow$REM transitions. Thus, our approach represents a simplifying approximation.  

The current study is further limited by the modest dataset size (24 fetal sheep) and the inherent ambiguity of the Intermediate state, which remains difficult to classify reliably. While our dataset was restricted to age-matched subjects between approximately 121 and 128 days of gestation, application of this framework to earlier or later gestational ages will require additional validation. These factors constrain the generalizability of our findings to broader populations, other developmental stages, and other species. 

Finally, while our study was conducted under appropriate animal ethics approval, translation to human studies introduces additional ethical and technical challenges. In particular, non-invasive fetal EEG recordings face substantial barriers related to extremely low signal-to-noise ratio.
% and Institutional Review Board compliance, which must be addressed before clinical application. 
Moreover, the proposed translational pathway (using the EEG sleep stager as a label engine, pretraining proxy stagers (e.g. fetal heart rate variability, ECG and / or ultrasound via weak/semi-supervised learning, and leveraging fMEG as a bridge to human EEG) has not yet been empirically validated, and its clinical impact on outcomes such as HDP remains to be demonstrated.

\subsection{Future Work}
Future work will explore more flexible non-linear spectral equalisation strategies to accommodate physiological differences. To address the poor classification of the Intermediate state, semi-supervised learning or data augmentation could be leveraged to improve the representation of ambiguous epochs. Alternatively, merging Intermediate with adjacent states may be considered if justified by physiological evidence, though this requires careful validation.  Another potential approach is to first train the model on REM and NREM data only, and then use prediction confidence as a metric to flag uncertain epochs as Intermediate, effectively treating the Intermediate state as a gray zone where the model is unable to classify with high certainty.  

Expanding the dataset to include a larger number of fetal sheep and a wider range of gestational ages will be essential for improving robustness and generalizability. Beyond animal models, validation against human fetal proxies such as fetal ECG, fMCG or Doppler Ultrasound will be important steps toward translational relevance. In this context, fMEG provides a promising non-invasive modality to facilitate cross-species transfer from sheep and neonatal EEG to human fetal EEG. Future studies should also explore large-scale weakly labeled datasets to enable proxy stager development from non-invasive signals, and longitudinal validation will be essential to determine whether state-aware monitoring improves early risk prediction of HDP.

In parallel, expanding multimodal recordings (e.g., EEG, ECG, EMG, EOG) in fetal sheep will help strengthen classification of ambiguous states and provide a richer foundation for proxy signal development. 
%Finally, future research should also integrate ethical considerations for human studies, including the design of non-invasive acquisition systems and compliance with regulatory frameworks, to bridge the gap between experimental fetal models and clinically feasible human monitoring.

\section{CONCLUSION}
This study introduced FetalSleepNet, the first deep learning framework for fEEG sleep staging, and proposed a label-efficient, domain-adaptive spectral equalisation method to adapt adult EEG for transfer learning. By reducing cross-domain mismatch, spectral equalisation improved training efficiency, stability, and generalization, particularly for REM/NREM classification, with minimal computational cost. While the Intermediate state remains challenging due to limited labeled examples, the proposed approach demonstrates robust performance under data scarcity and cross-domain conditions, providing a practical foundation for automated and scalable fetal sleep staging, and establishing the EEG sleep stager as a label engine to generate reliable annotations for downstream proxy stagers. 
% This work has strong potential for translation to human fetal monitoring, enabling state-specific physiological analyses and proxy model development from non-invasive signals.
This work has strong potential for translation to human fetal monitoring, enabling state-specific physiological analyses and weak/semi-supervised pretraining of proxy stagers from non-invasive signals such as Doppler ultrasound and fECG. As a brain-based bridge, fMEG further supports transfer from sheep and neonatal EEG to human fetal EEG, paving the way toward clinical applications.
Moreover, owing to its lightweight and computationally efficient design, FetalSleepNet is well suited for low-power, real-time, and potentially wearable fetal monitoring systems, supporting longitudinal deployment in large clinical cohorts. 
% Future work will integrate additional domain adaptation techniques, validate on non-invasive human fEEG, and extend the framework to multimodal physiological sensing for comprehensive fetal neurodevelopment assessment.
Future work will integrate additional domain adaptation techniques, and extend the framework to multimodal physiological sensing for comprehensive fetal neurodevelopment assessment. Importantly, prospective studies will be required to test whether such state-aware monitoring improves early risk prediction of fetal and maternal health complications.
%The findings support the development of automated, objective fetal sleep monitoring tools. 
%Future work will explore transfer learning from adult sleep EEG models to improve fetal sleep state classification, particularly in scenarios with limited labeled fetal data.
\bibliographystyle{IEEEtran}
\bibliography{Reference}

\end{document}